\documentclass[conference]{IEEEtran}

\usepackage{algorithm}
\usepackage{algorithmic}

\IEEEoverridecommandlockouts
\usepackage{cite}
\usepackage{amsmath,amssymb,amsfonts}
\usepackage{algorithmic}
\usepackage{graphicx}
\usepackage{textcomp}
\usepackage{xcolor}
\def\BibTeX{{\rm B\kern-.05em{\sc i\kern-.025em b}\kern-.08em
    T\kern-.1667em\lower.7ex\hbox{E}\kern-.125emX}}

\usepackage{graphicx}
\usepackage{subcaption}
\usepackage{tabularx}

\begin{document}

\title{DNN-Driven Proactive UAV Trajectory and Beam Steering for Blockage-Resilient Communications
		\vspace{-0.3em}}
	
	\author{{Abdallah S. Ghazy, Hussein~A.~Ammar, James~Bayes, and Francois Chan}	
		
		\\
		Dept. of Elec. and Comp. Eng., Royal Military College of Canada, Kingston, ON, Canada \\
		Emails: abdallah.ghazy@rmc-cmr.ca, hussein.ammar@rmc.ca, james.bayes@rmc-cmr.ca,  chan-f@rmc.ca
		\vspace{-1em}
	}
	
	\maketitle
		
	\begin{abstract}
		This paper proposes Autonomous UAV Communication (AutUAVCom), a proactive communication system that jointly adapts the flight trajectory and RF beam steering to minimize  UAV energy consumption while satisfying secrecy and reliability requirements for rescue teams operating in dynamic, cluttered, and GPS-denied environments. Unlike existing model-based proactive approaches, AutUAVCom employs a deep neural network-based Tracking and Prediction (T\&P) module that combines onboard sensing with spatiotemporal and body-orientation information to track users, localize blocking objects, and predict future trajectories and potential beam blockages. Building on the T\&P module, we formulate a constrained optimization problem and develop a two-step iterative numerical approach. Monte Carlo simulations show that, relative to a reactive system, AutUAVCom achieves $10\%$ and $19\%$ gains in energy efficiency and channel secrecy, respectively, while reducing the outage probability by $85\%$, under a mean blockage arrival interval of $120$ seconds and a mean blockage duration of $20$ seconds.
	\end{abstract}
	
	
	\vspace{-0.2cm}
	\section{Introduction}
Unmanned Aerial Vehicles (UAVs) provide flexible and rapidly deployable wireless relaying in challenging environments
. However, battery constraints, time-varying channels, user mobility, navigation challenges, and eavesdropping risks require intelligent real-time control rather than offline trajectory design \cite{UAV_Survey_2019}. Model Predictive Control (MPC) addresses these dynamics by repeatedly solving finite-horizon constrained optimization problems \cite{MPC_Book_2025}. Recent MPC frameworks optimize UAV trajectory, power, and security \cite{MPC_Solar_2021, MPC_Security_2025}, but their performance depends strongly on accurate state estimation and mobility prediction.

Existing MPC frameworks mainly rely on simplified statistical or model-based mobility assumptions. Tuan \textit{et al.} \cite{MPC_Solar_2021} used statistical air-to-ground channel models without future channel state information (CSI), while Sheng \textit{et al.} \cite{MPC_Security_2025} employed a Gauss-Markov mobility (GMM) model for trajectory prediction under location uncertainty. Similarly, Huang \textit{et al.} \cite{UAV_FullDuplex_2025} modeled user motion using random mobility with predicted CSI errors, Jin \textit{et al.} \cite{WNCS_2026} predicted control states probabilistically, and Savkin \textit{et al.} \cite{3D_Securing_2020} assumed road-constrained constant-velocity motion. These approaches face four key limitations: predefined stochastic models may mismatch real pedestrian or team motion, repeatedly performing online model-based trajectory prediction and control can incur computational and propulsion overhead,  environmental disturbances such as wind can cause displacement and beam misalignment, and GPS/CSI-based localization assumes accurate position information, limiting operation in GPS-denied environments that require onboard perception.

To address GPS-denied tracking, we previously developed a vision-inertial tracking framework \cite{Ghazy_2026_UAV}. Building on this framework, we propose Autonomous UAV Communication (AutUAVCom), a proactive communication system that jointly adapts UAV trajectory and RF beam steering to mitigate coverage loss and anticipate future blockages, enabling energy-efficient, reliable, and secure communication for rescue teams in dynamic, cluttered, and GPS-denied environments.  Unlike model-based approaches \cite{MPC_Solar_2021, MPC_Security_2025, UAV_FullDuplex_2025, WNCS_2026, 3D_Securing_2020}, AutUAVCom learns future user positions from historical spatiotemporal and body-orientation observations via offline deep neural network (DNN) training. Once trained, it directly performs online inference from new observations, without retraining or repeatedly solving a mobility model. In this paper, we develop a DNN-based Tracking and Prediction (T\&P) module that uses onboard sensing to track users, localize blocking objects, and predict future user trajectories from historical spatiotemporal and body-orientation information. We formulate a constrained optimization problem that minimizes UAV energy consumption while satisfying secrecy and reliability requirements, and develop a two-step iterative numerical approach. Comprehensive Monte Carlo simulations model blockage events as Poisson processes and demonstrate the effectiveness of the proposed approach. The remainder of this paper is organized as follows. Section \ref{sec:system_model} presents the system model. Section \ref{sec:mpc} introduces the T\&P model and problem formulation. Section \ref{Numerical_Results} presents the numerical results. Finally, Section \ref{Conclusion} concludes the paper.

\vspace{-0.25cm}

	\begin{figure*}[t]
		\centering
		\includegraphics[width = .95\linewidth]{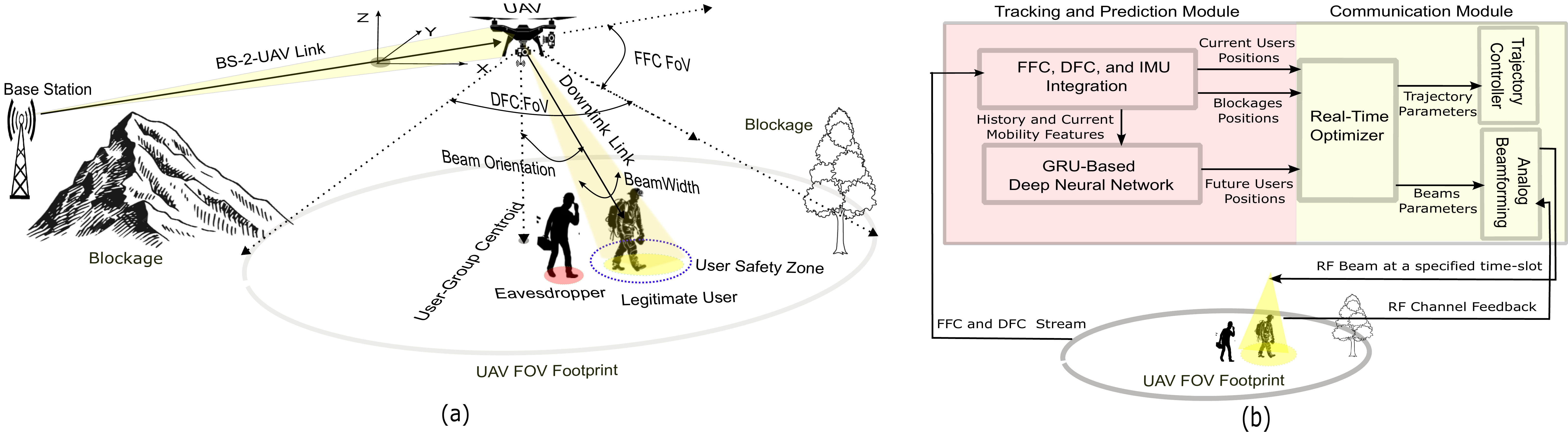}
		\caption{ The proposed AutUAVCom system; (a) Considered network topology, and (b) A system block diagram.}
		\label{System_Model}
		\vspace{-0.6cm}
	\end{figure*}
	
\section{System Model} 
\label{sec:system_model}
\subsection{System Overview}
As illustrated in Fig.~1(a), a UAV serves a rescue team in a cluttered, dynamic, and GPS-denied environment. Using onboard sensing, the UAV estimates user and blocking-object states and anticipates future beam blockages; its coverage region is determined by its altitude and antenna/cameras field of view (FOV). It jointly adapts its trajectory and RF beam steering to maintain reliable coverage of legitimate users (Rxs) while avoiding locations occupied by potential eavesdroppers (Exs). To improve energy efficiency and security, its trajectory avoids unnecessary movement while accounting for Rx safety zones, wind-induced displacement, and attitude fluctuations that may cause beam misalignment and information leakage toward Exs.

As illustrated in Fig.~\ref{System_Model}(b), AutUAVCom implements a closed-loop tracking-prediction-optimization-communication process comprising two integrated modules: the T\&P Module and the Communication Module. The T\&P Module integrates data from a downward-facing camera (DFC), forward-facing camera (FFC), and Inertial Measurement Unit (IMU) to estimate current user positions and localize blocking objects. Historical positions and mobility features, including speed and body orientation, are then provided to a Gated Recurrent Unit (GRU)-based DNN to predict future user positions over the operational horizon. The Communication Module combines the current user and blocking-object states, predicted user trajectories, and continuous RF channel feedback to optimize the UAV trajectory and RF beam parameters. The resulting trajectory commands are sent to the trajectory controller, while the beam parameters are provided to the analog beamforming unit. 
\vspace{-0.15cm}
\subsection{Channel Model}
\subsubsection{\textbf{Path Loss}} 
With the UAV (Tx) at the origin, the BS, $k$th Rx, and its associated Ex are located at $(x_b,y_b,z_b)$, $(x_k,y_k,z_k)$, and $(x_e,y_e,z_e)$, respectively. A safety zone of radius $d_s=\sqrt{(x_k-x_e)^2+(y_k-y_e)^2}$ separates each Rx from its eavesdropper. Assuming a LoS-dominant BS-to-UAV backhaul, its path loss in dB is~\cite{2014_PathLoss}
	
	\begin{equation}
		g_{b|{\rm dB}} = - 20\log_{10}\left(\frac{4\pi f_c d_b}{c}\right) -\eta_b^{\mathrm{LoS}},
		\label{eq:gain_backhaul}
	\end{equation}
	where $f_c$ is the carrier frequency, $c$ is the speed of light, $d_b=\sqrt{x_b^2+y_b^2+z_b^2}$, and $\eta_b^{\mathrm{LoS}}$ is the excess LoS loss. The UAV-$k^{th}$ user (access) link follows an air-to-ground model that accounts for the LoS probability and the corresponding LoS/NLoS path losses, which is written in dB as~\cite{2014_PathLoss}:
	
	\vspace{-0.4em}
	\begin{align}
		g_{k|{\rm dB}}^l &= -\frac{\eta_{\mathrm{LoS}}-\eta_{\mathrm{NLoS}}}{1+a\exp\left(-b\left(\frac{180}{\pi}\psi_k-a\right)\right)} - 20\log_{10}(d_k) 
		\nonumber\\
		&\ \ \ - 20\log_{10}\left(\frac{4\pi f_c}{c}\right) - \eta_{\mathrm{NLoS}},
		\label{eq:gain_access}
	\end{align}
	where $d_k=\sqrt{x_k^2+y_k^2+z_k^2}$, $\psi_k=\sin^{-1}(|z_k|/d_k)$ is the elevation angle, $\eta_{\mathrm{LoS}}$ and $\eta_{\mathrm{NLoS}}$ are the excess losses, and $a$ and $b$ are environment-dependent empirical parameters that characterize the surrounding obstacle density and geometry ~\cite{2014_PathLoss}. 
	
	\subsubsection{\textbf{Signal Blockage}}
	To characterize transient blockages caused by physical blocking objects, such as trees, blockage arrivals are modeled according to a Poisson process with rate~$\lambda_b$. Accordingly, the inter-arrival time between consecutive blockage events follows an exponential distribution with probability density function (PDF) as $
	f_{\lambda_b}(t) = \lambda_b e^{-\lambda_b t}, \,t\geq0,$
	where $\lambda_b$ [events/s] is the average blockage arrival rate and $T_b=1/\lambda_b$ is the mean inter-blockage time~\cite{Gallager2013}. The duration of an individual blockage event is independently modeled by an exponential distribution with PDF as $
	f_{\lambda_d}( \tau) = \lambda_d e^{-\lambda_d  \tau}, \, \tau\geq0,$
	where $\lambda_d$ [1/s] is the blockage clearance rate and $T_d=1/\lambda_d$ represents the mean blockage duration~\cite{Gallager2013}. The resulting blockage state is represented by the binary variable $S_b(t)\in\{0,1\}$, where $S_b(t)=1$ indicates the presence of an RF blockage and $S_b(t)=0$ otherwise.

	\subsubsection{\textbf{Pointing Errors}}
	Wind-induced turbulence displaces and disorients the UAV. Since the backhaul employs a wide beam and has no nearby eavesdroppers, pointing errors are neglected for the backhaul but are considered on the access link, where a narrow beam is employed for communication efficiency and security. Following the experimentally validated statistical model of~\cite{Ghazy_UAV_PDFs,2013_Hovering_PDF}, UAV displacement is modeled as independent Gaussian variables along each axis, while beam disorientation follows a Cox-Munk polar-angle distribution with a truncated-Gaussian azimuth angle. The resulting PDF of the pointing-error gain $g_k^p$ is~\cite{Farid_Steve}
	\vspace{-0.4em}
	\begin{equation}
		f_{g_k^p}(g_k^p) = \dfrac{\xi_k^2}{A_k^{\xi_k^2}} (g_k^p)^{(\xi_k^2-1)}, \qquad 0\leq g_k^p \leq A_k,
		\label{PDF_Pointing_Error}
	\end{equation}
	where $\xi_k^2 = \frac{\sqrt{\pi}\operatorname{erf}(\epsilon_k)w_k^2}{8\Lambda_k^2\epsilon_k\exp(-\epsilon_k^2)}$, $A_k=[\operatorname{erf}(\epsilon_k)]^2$, $\epsilon_k=\frac{\sqrt{\pi}a_k}{\sqrt{2}w_k}$, and $\Lambda_k^2$, $a_k$, and $w_k$ are the jitter variance, receiver aperture radius, and beamwidth of the $k$th Rx, respectively. Equation~\eqref{PDF_Pointing_Error} is accurate for $w_k>6a_k$~\cite{Farid_Steve}.

	\subsubsection{\textbf{SNR, Outage, and Secrecy}}
	Considering the effects of path loss, transient blockages, and pointing errors, the 
	SNR of the UAV- $k^{\text{th}}$ Rx link is expressed as
	\begin{equation}
		\gamma_k(t) = \frac{g_u g_k^p(t) g_k^l(t) p_k(t)}{\sigma_k^2} (1-S_b(t)),
		\label{eq:SNR}
	\end{equation}
	where $g_u$ is the UAV antenna gain, $p_k$ is the transmit power, and $\sigma_k^2$ is the noise power at the $k$th Rx.  The corresponding channel capacity is
	$c_k (t) = \log_2(1+\gamma_k(t))$ bits/channel use
	~\cite{2005_Goldsmith_Wireless}. The outage probability for a capacity threshold $c_{\mathrm{th}}$ is given by
	\begin{equation}
		F_{k}(c_{\mathrm{th}})= \int_{0}^{c_{\mathrm{th}}} f_{c_k}(c)\,\mathrm{d}c,
		\label{P_out}
	\end{equation}
	where $f_{c_k}(c)$ denotes the PDF of the instantaneous channel capacity. Finally, physical-layer security is quantified by the channel secrecy rate (CSR)~\cite{Ghazy_UAV_PDFs}, defined as $
	\mathrm{CSR}_k = \max\left(0, \hat{c}_k - \hat{c}_{k'}\right),$ 
	where $\hat{c}_k$ and $\hat{c}_{k'}$ denote the average channel capacities of the $k^{th}$ Rx and its associated Ex, respectively, assuming the worst-case scenario in which the Ex is located at a distance $d_s$ from its associated Rx.

	\subsubsection{\textbf{Energy Consumption}}
	The UAV's total mission energy combines propulsion energy $E_{\rm pro}$, hovering energy $E_{\rm hov}$, and communication energy $E_{\rm com}$,  expressed as $
	E_{tot}=E_{\rm pro}+E_{\rm hov}+E_{\rm com}\,$,  ~\cite{2019_Pwer_Hovering_Model}.
	
	
	\section{Prediction and Problem Formulation}
	\label{sec:mpc}

	\begin{figure}[t]
		\centering
		\includegraphics[width = .9999\linewidth]{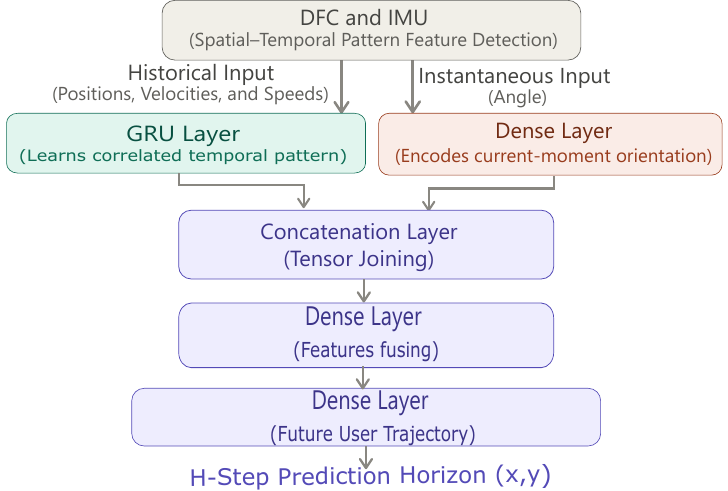}
		\caption{  DNN architecture of T\&P module. }
		\label{Two_Branch_Gru_Architecture}
		\vspace{-0.4cm}
	\end{figure}

	\subsection{Tracking and Prediction (T\&P) Module}
    Figure ~\ref{Two_Branch_Gru_Architecture} details the GRU-based T\&P architecture.
	
	\subsubsection{Tracking Unit}
	At each time instant, $t$, the onboard DFC tracks users, while the FFC localizes blocking objects, the IMU measures the UAV's linear acceleration and angular velocity to account for UAV motion and attitude changes. The tracking unit fuses these measurements to estimate the UAV position in real time. Together with the calibration parameters of the DFC and FFC models, this enables the image observations to be reconstructed in 3D and transformed into the UAV frame, yielding the relative locations of users and blocking objects. The DFC also estimates user velocity and speed from consecutive frames and infers the users' moving directions from their body orientations~\cite{body_orientation_2009}. Coordinate projections follow our vision-inertial tracking framework in \cite{Ghazy_2026_UAV}. The mobility state of user $k$ at time $t$ is formulated as
	\begin{equation}
		\mathbf{s}_k[t]=
		\left[
		x_k[t],\,y_k[t],\,\nu_{x,k}[t],\,\nu_{y,k}[t],\,\nu_k[t],\,\alpha_k[t]
		\right]^{\mathrm T},
		\label{eq:state_vector}
	\end{equation}
	where $(x_k[t],y_k[t])$ is the position, $\nu_{x,k}[t]=dx_k/dt$ and $\nu_{y,k}[t]=dy_k/dt$ are the velocity components, $\nu_k[t]=\sqrt{\nu_{x,k}^2[t]+\nu_{y,k}^2[t]}$ is the moving speed, and $\alpha_k[t]\approx\operatorname{atan2}\left(y_k[t{+}1], x_k[t{+}1]\right)$ is the estimated body orientation.
	
	\subsubsection{Prediction Unit}
	The prediction unit learns a nonlinear mapping from the historical mobility sequence and the current body orientation to the future trajectory. As illustrated in Fig.~\ref{Two_Branch_Gru_Architecture}, the prediction unit consists of historical and instantaneous branches. The historical branch processes the spatial-temporal mobility features, including positions, velocities, and speeds, over an observation window of length $L$ using a single GRU layer, and its input is formulated as
	\begin{equation}
		\mathbf{X}_k[t]=
		\left[
		\bar{\mathbf{s}}_k[t-L+1],\ldots,\bar{\mathbf{s}}_k[t]
		\right]^{\mathrm T},
		\label{eq:gru_input}
	\end{equation}
	where $\bar{\mathbf{s}}_k[t]$ matches $\mathbf{s}_k[t]$ but omits body orientation, which is instead provided to the instantaneous branch. The latter processes the current orientation using a Dense layer with a ReLU activation function. The outputs of the two branches are then concatenated and passed through a Dense layer  for feature fusion. Finally, a Dense layer maps the fused features to the predicted $(x,y)$ positions over the next $H$ time-steps.

\subsubsection{Dataset and Training}
The prediction unit is trained offline in a supervised manner. Due to the lack of real-world datasets characterizing the mobility patterns of rescue teams, we generate the training dataset using an analytical joint mobility model based on the Reference Point Group Mobility (RPGM) model, where the reference-point trajectory follows a GMM. The proposed prediction framework is not restricted to RPGM-GMM dataset and can be applied to datasets exhibiting spatial and temporal correlations. Given $N_{\rm}$ training samples, the optimal prediction parameters, $\boldsymbol{\omega}^{*}$, are obtained by minimizing the mean square error (MSE) loss:
\begin{equation}
		\vspace{-0.4em}
		\boldsymbol{\omega}^{*}
		=
		\underset{\boldsymbol{\omega}}{\arg\min}
		\;
		\frac{1}{N}
		\sum_{n=1}^{N}
		\left\|
		{\mathbf{p}_k}^{(n)}
		-
		f_{\rm p}
		\left(
		\mathbf{X}^{(n)}_k[t],
		\alpha_k^{(n)}[t];
		\boldsymbol{\omega}
		\right)
		\right\|_2^2,
		\label{eq:mse_loss}
	\end{equation}
where $\boldsymbol{\omega}$ denotes the trainable parameters of the DNN, $\mathbf{p}_k$  denotes ground-truth trajectories of $k^{th}$ Rx, $f_{p}(\cdot)$  denotes the parameterized mapping function, and $\chi^{(n)}$ denotes the  $n^{th }$ training sample of $\chi$. The optimization is performed using the Adam optimizer with a small learning rate (e.g., $10^{-4}$). The dataset is split into training, validation, and test subsets. Early stopping based on the validation loss, with restoration of the best-performing weights, is used to prevent overfitting, while all inputs are min-max normalized based on the training statistics. The held-out test subset provides an unbiased performance estimate. During online operation, the trained model predicts the next $H$ positions of $k^{th}$ Rx, $\hat{\mathbf{p}}_k[t+1:t+H]$, as
	\begin{equation}
		\hat{\mathbf{p}}_k[t+1:t+H]
		=
		f_{\rm p}
		\left(
		\mathbf{X}_k[t],
		\alpha_k[t];
		\boldsymbol{\omega}^{*}
		\right)\,.
		\label{eq:gru_predict}
	\end{equation}
The Prediction unit uses  the predicted user trajectories to infer the $H$ future trajectories of the user-group centroid, $\hat{\mathbf{p}}_c[t+1:t+H]$. It also predicts potential blockages by identifying intersections between the predicted user trajectories and the blockage locations obtained by tracking unit. 
\vspace{-0.1cm}

\begin{figure*}[!t]
\setlength{\belowdisplayskip}{-6pt}
		\begin{align}
			\underset{
				\substack{
					\{\mathbf{p}_u[t+i|t],\,w_k[t+i|t]\}_{i=0}^{H-1}
				}}
				{\mathrm{minimize}}
				\,
				&E_{\mathrm{tot}}[t:t+H|t]
				=
				\sum_{i=0}^{H-1}
				\left(
				E_{\mathrm{pro}}(\mathbf{p}_u[t+i|t])
				+
				E_{\mathrm{hov}}(\mathbf{p}_u[t+i|t])
				+
				E_{\mathrm{com}}(\mathbf{p}_u[t+i|t],w_k[t+i|t])
				\right)
				\label{eq:Pro_Form_MPC}
			\end{align}
			\setcounter{equation}{\value{equation}}
			\noindent
			\begin{minipage}[t]{0.49\textwidth}
				\begin{align}
					\mathrm{subject~to:}\,\,\,&
					\mathrm{CSR}_k[t+i|t]\geq \mathrm{CSR}_{\mathrm{th}},
					\quad \forall k,
					\tag{\theequation a}\label{eq:CSR}\\
					&
					F_k[t+i|t]\leq F_{\mathrm{th}}, \quad \forall k,
					\tag{\theequation b}\label{eq:out}\\
					&
					c_k[t+i|t]\leq c_b[t+i|t], \quad \forall k,
					\tag{\theequation c}\label{eq:cap}\\
					&
					0\leq p_k[t+i|t]\leq P_u^{\max}, \quad \forall k,
					\tag{\theequation d}\label{eq:pow1}\\
					&
					0\leq p_b[t+i|t]\leq P_b^{\max},
					\tag{\theequation e}\label{eq:pow2}\\
					&
					\mathbf{p}_{u}[t+i|t]
					\notin
					\bigcup_{q=1}^{Q}
					\mathcal{B}_{q}[t+i|t],
					\tag{\theequation f}\label{eq:blocking}
				\end{align}
			\end{minipage}
			\hfill
			\begin{minipage}[t]{0.49\textwidth}
				\begin{align}
					&
					\left\|
				\mathbf{p}_u[t+i|t]
					-
					\hat{\mathbf{p}}_c[t+i|t]
					\right\|
					\leq R_{\mathrm{FOV}}-R_{\max},
					\tag{\theequation g}\label{eq:FOV}\\
					&
					\max\left\{
					\frac{R_{\max}+\Delta}
					{\tan(\theta_{\mathrm{FOV}}/2)},
					h_{\min}^{b}
					\right\}
					\leq z_u[t+i|t]\leq h_{\max},
					\tag{\theequation h}\label{eq:height}\\
					&
					0\leq \theta_k[t+i|t]
					\leq \frac{\theta_{\mathrm{FOV}}}{2}, \quad \forall k,
					\tag{\theequation i}\label{eq:theta}\\
					&
					0\leq \phi_k[t+i|t]\leq 2\pi, \quad \forall k,
					\tag{\theequation j}\label{eq:phi}\\
					&
					w_{\min}\leq w_k[t+i|t]\leq w_{\max}, \quad \forall k.
					\tag{\theequation k}\label{eq:bw}
				\end{align}
			\end{minipage}
			
			\vspace{1mm}
			\hrulefill
			\vspace{-0.2cm}
		\end{figure*}

	\subsection{Communication Module}

\vspace{-0.2cm}
\subsubsection{Problem Formulation}
The AutUAVCom system should minimize UAV  energy consumption while maximizing CSR and reliability. However, these objectives are inherently coupled: energy-efficient trajectories may reduce user tracking accuracy, whereas CSR and reliability require appropriate user coverage. Similarly, wider beams improve coverage robustness at the cost of increased leakage, while narrower beams improve CSR but reduce coverage tolerance. Wind-induced fluctuations further affect these trade-offs. We therefore formulate  a constrained single-objective optimization, as given in  \eqref{eq:Pro_Form_MPC}, where total UAV energy consumption is minimized subject to minimum CSR and maximum outage-probability requirements. The optimization determines the UAV trajectory $\mathbf{p}_u[t+i]$ and RF beamwidth $w_k[t+i]$ over the next $H$ time steps, $i=\{0,\ldots,H-1\}$, subject to  \eqref{eq:CSR}-\eqref{eq:bw}. Specifically, constraint~\eqref{eq:CSR} guarantees that the CSR   remains above the required threshold, $\text{CSR}_{\mathrm{th}}$, and  constraint~\eqref{eq:out} limits the outage probability to $F_{\mathrm{th}}$. Constraint~\eqref{eq:cap} ensures that the capacity $c_k$ does not exceed the available backhaul capacity $c_b$. Constraints~\eqref{eq:pow1} and~\eqref{eq:pow2} bound the access and backhaul transmit powers by their respective maximum allowable values, i.e., $P_u^{\max}$ and $P_b^{\max}$. Constraint~\eqref{eq:blocking} prevents the UAV from entering the predicted blockage regions $\mathcal{B}_{q}[t+i|t]$,	where $\mathcal{B}_{q}$ is the estimated blockage zone of the $q^{th}$ blocking object. Constraint~\eqref{eq:FOV} maintains the UAV position within the effective RF antenna/DFC FOV, where  $R_{\mathrm{FOV}}$ is the radius of the RF coverage region, and $R_{\max}$ is the maximum radius of the user group, where $R_{\mathrm{FOV}}>R_{\max}$. Constraint~\eqref{eq:height} restricts the UAV altitude between the minimum height required to provide coverage over the user group with radius $R_{\max}$ and the maximum allowable height $h_{\max}$. Here, $h_{\min}^{b}$ denotes the minimum UAV height required to maintain visibility of the BS, $\theta_{\mathrm{FOV}}$ is the RF antenna/DFC angle FOV, and $\Delta=R_{\mathrm{FOV}}-R_{\max}$ represents the coverage margin. Constraints~\eqref{eq:theta} and~\eqref{eq:phi} constrain the polar and azimuth angles of the  RF beam directed to  $k^{\text{th }}$ Rx to their physically feasible ranges. Finally, constraint~\eqref{eq:bw} restricts the RF beamwidth to $[w_{\min},w_{\max}]$, thereby balancing beam directivity, pointing-error sensitivity, and spatial selectivity.

\begin{algorithm}[t]
\caption{Pseudocode of the ET-BSMO Algorithm}
\label{alg:ET_BSMO}
\begin{algorithmic}[1]
\REQUIRE Current UAV state $\mathbf{p}_u[t]$, sensors measurements, QoS feedback,
prediction horizon $H$, threshold $M_{\rm th}$
\ENSURE Suboptimal UAV trajectory $\{\mathbf{p}_u^*[t+i]\}_{i=0}^{H-1}$

\STATE Compute user visibility margins $m_k[t]$ and worst-case margin $M[t]$.
\STATE Evaluate QoS feasibility indicator $Q[t]$.
\IF{$M[t]\geq M_{\rm th}$ and $Q[t]=1$}
    \STATE Maintain $\mathbf{p}_u^*[t+i]=\mathbf{p}_u[t]$ for $i=\{0,\ldots,H-1\}$;
    update RF beam direction and width.
\ELSE
    \FOR{$i=\{0,\ldots,H-1\}$}
        \STATE Estimate user-group centroid $\hat{\mathbf p}_c[t+i|t]$.
        \STATE Compute dominant direction $\hat{\mathbf d}[t+i|t]$
        using centroid, prediction-error, and detouring directions.
        \STATE Initialize search bounds for displacement $d$ and angular
        adjustments $(\vartheta_u,\varphi_u)$.
        \WHILE{search has not converged}
            \STATE Evaluate the midpoint candidate for FOV, blockage, and QoS feasibility.
            \STATE Update the corresponding binary-search bounds.
        \ENDWHILE
        \STATE Obtain $d^*[t+i|t]$ and $\hat{\mathbf d}^*[t+i|t]$.
        \STATE Update
        $\mathbf{p}_u^*[t+i+1]
        =\mathbf{p}_u[t+i]
        +d^*[t+i|t]\hat{\mathbf d}^*[t+i|t]$.
    \ENDFOR
\ENDIF
\RETURN $\{\mathbf{p}_u^*[t+i]\}_{i=0}^{H-1}$
 \STATE Apply only $\mathbf{p}_u^*[t+1]$
  \STATE Go to line No. 1
\end{algorithmic}
\end{algorithm}

\subsubsection{Event-Triggered Binary-Search Motion Optimization (ET-BSMO)}
Problem~\eqref{eq:Pro_Form_MPC} is nonlinear and nonconvex because the achievable rates depend nonlinearly on the UAV position through distance-dependent path loss and time-varying user locations, while propulsion energy couples consecutive UAV states over the prediction horizon. Since a closed-form solution is generally intractable, we propose ET-BSMO, a computationally efficient iterative approach. Algorithm~\ref{alg:ET_BSMO} presents its pseudocode, combining event-triggered adaptation with proactive motion optimization over an $H$-step prediction horizon. To reduce unnecessary maneuvers, the algorithm optimizes the horizontal trajectory at a fixed altitude, allowing vertical adjustments only to avoid obstructing objects.  The algorithm is described as follows.

\vspace{-0.05cm}
\paragraph{Event-Triggered Adaptation}
At time $t$, the UAV determines whether physical repositioning is required using sensors measurements and QoS feedback. Let
$\mathbf{p}^{c}_{k}[t]=[u_k[t],v_k[t]]^{\top}$ denote the pixel coordinates of user $k$ in the DFC frame. The normalized coordinates are
\begin{equation}
\tilde{u}_{k}[t]=\frac{2u_k[t]}{u_{\max}}-1,\qquad
\tilde{v}_{k}[t]=\frac{2v_k[t]}{v_{\max}}-1,
\end{equation}
where $u_{\max}$ and $v_{\max}$ are the image width and height, respectively. The corresponding FOV margin and worst-case margin  are computed as
\begin{equation}
m_k[t]=1-\max\left(|\tilde{u}_k[t]|,|\tilde{v}_k[t]|\right),\,
M[t]=\min m_k[t].
\end{equation}
Let $Q[t]=1$ when all users satisfy the required QoS constraints and $Q[t]=0$ otherwise. These quantities form the event trigger. If $(M[t]\geq M_{\rm th})\wedge(Q[t]=1),$ where $\wedge$ denotes the logical AND operation, 
the UAV remains stationary and only updates its RF beam direction and width to accommodate minor user movements. This hovering policy avoids unnecessary propulsion energy consumption. Otherwise, a motion action is activated over the next $H$ time steps.

\vspace{-0.1cm}
\paragraph{Proactive Motion Policy}
When repositioning is required, the UAV first estimates/predicts user-group centroid and determines a dominant motion direction as
\begin{equation}
\hat{\mathbf d}[t+i|t]=
\frac{
\hat{\mathbf d}_o[t+i|t]
+\hat{\mathbf d}_e[t+i|t]
+\hat{\mathbf d}_d[t+i|t]
}{
\left\|
\hat{\mathbf d}_o[t+i|t]
+\hat{\mathbf d}_e[t+i|t]
+\hat{\mathbf d}_d[t+i|t]
\right\|}.
\label{eq:dominant_direction}
\end{equation}
The nominal direction toward the user-group centroid is
\begin{equation}
\hat{\mathbf d}_o[t+i|t]=
\frac{
\hat{\mathbf p}_c[t+i|t]-\mathbf p_u[t+i|t]
}{
\left\|
\hat{\mathbf p}_c[t+i|t]-\mathbf p_u[t+i|t]
\right\|
},
\label{eq:nominal_direction}
\end{equation}
where $\hat{\mathbf p}_c[t+i|t]$ is the predicted user-group centroid. The prediction-error correction is given by
\begin{equation}
\hat{\mathbf d}_e[t+i|t]=
\frac{\mathbf e_g[t+i|t]}
{\|\mathbf e_g[t+i|t]\|},
\qquad
\|\mathbf e_g[t+i|t]\|>0,
\label{eq:error_direction}
\end{equation}
where $\mathbf e_g[t+i|t]=\mathbf{p}_c[t+i|t]-\hat{\mathbf{p}}_c[t+i|t]$ denotes the residual between the actual and predicted group centroids. When a blockage is predicted, $\hat{\mathbf d}_m[t+i|t]$ provides the required obstacle-avoidance adjustment. Specifically,
\begin{equation}
\hat{\mathbf d}_d[t+i|t]=
\begin{bmatrix}
\sin(\theta_u+\vartheta_u)\cos(\phi_u+\varphi_u)\\
\sin(\theta_u+\vartheta_u)\sin(\phi_u+\varphi_u)\\
\cos(\theta_u+\vartheta_u)
\end{bmatrix},
\end{equation}
where $(\theta_u,\phi_u)$ describes the direction of $\hat{\mathbf d_o}^{*}[t+i|t]+\hat{\mathbf d_e}[t+i|t]$ and $(\vartheta_u,\varphi_u)$ are geometry-dependent angular adjustments selected to avoid predicted blockage while preserving the FOV and QoS constraints. The UAV is then restricted to move along the resulting direction, reducing the original search space to the displacement and detouring variables. Its position is recursively updated as
\begin{equation}
\mathbf p_u^{*}[t+i+1]
=
\mathbf p_u[t+i]
+d^{*}[t+i|t]\hat{\mathbf d}^{*}[t+i|t],
\end{equation}
where $\hat{\mathbf d}^{*}[t+i|t]$ is the selected feasible direction
and $d^{*}[t+i|t]$ is the corresponding displacement. The displacement and angular adjustments are obtained by exploiting the monotonic feasibility of candidate positions along the selected direction and applying binary search over
\begin{equation}
\begin{split}
&0<\vartheta_u[t+i|t]\leq\frac{\theta_{\rm FOV}}{2},
\qquad
0<\varphi_u[t+i|t]\leq2\pi, \\&
0<d[t+i|t]\leq
\left\|\hat{\mathbf p_c}[t+i|t]-\mathbf p_u[t+i|t]\right\|.
\end{split}
\end{equation}
 At each iteration, the midpoint candidate is evaluated against the FOV, blockage, and QoS constraints, and the search interval is updated until convergence. Repeating this procedure over the $H$-step horizon yields the suboptimal trajectory
$\{\mathbf p_u^{*}[t+i+1]\}_{i=0}^{H-1}$. The resulting computational complexity is
$\mathcal{O}(HJ\mathcal{O}_c)$, where $\mathcal{O}_c$ is the cost of one feasibility evaluation and $J$ is the total number of binary-search iterations over displacement, polar angle, and azimuth angle.  
\vspace{-0.1cm}
\begin{figure*}[t]
			\centering
			\begin{subfigure}{0.32\textwidth}
				\centering
				\includegraphics[width=0.9\linewidth]{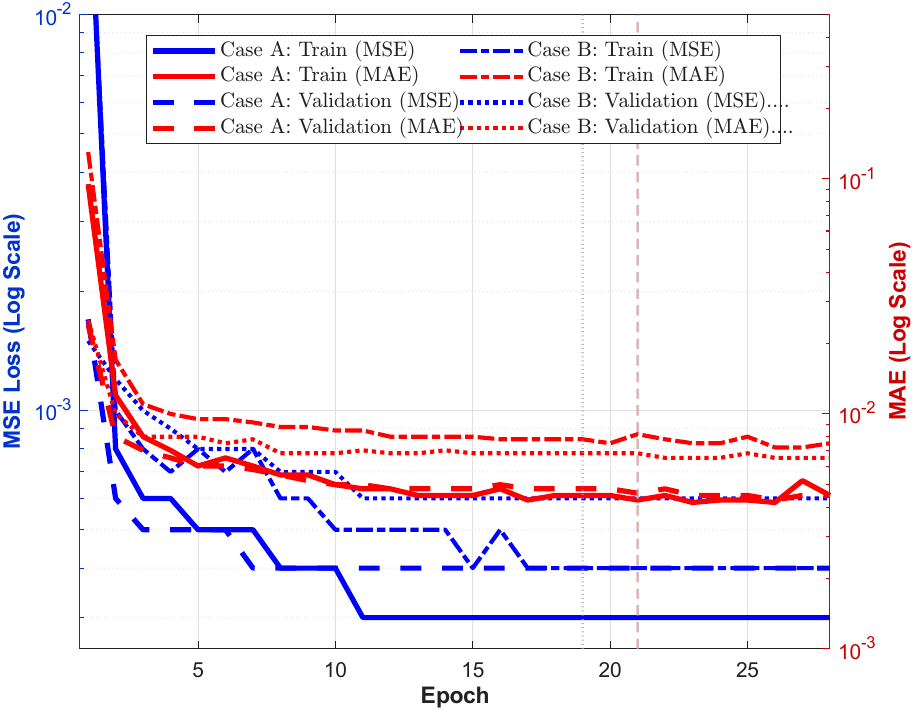}
				\caption{}
				\label{fig:MSE}
			\end{subfigure}
			\hfill
			\begin{subfigure}{0.32\textwidth}
				\centering
				\includegraphics[width=0.9\linewidth]{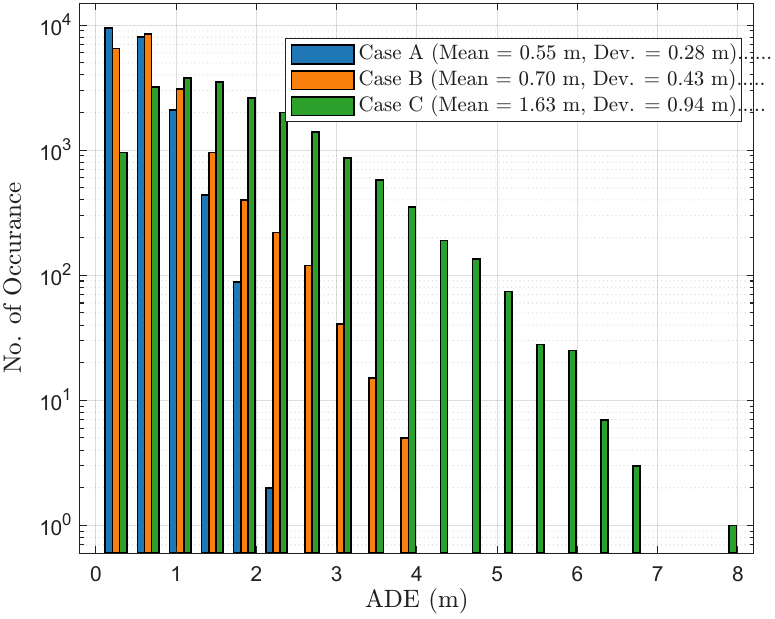}
				\caption{}
				\label{fig:ADE}
			\end{subfigure}
			\hfill
			\begin{subfigure}{0.32\textwidth}
				\centering
				\includegraphics[width=0.9\linewidth]{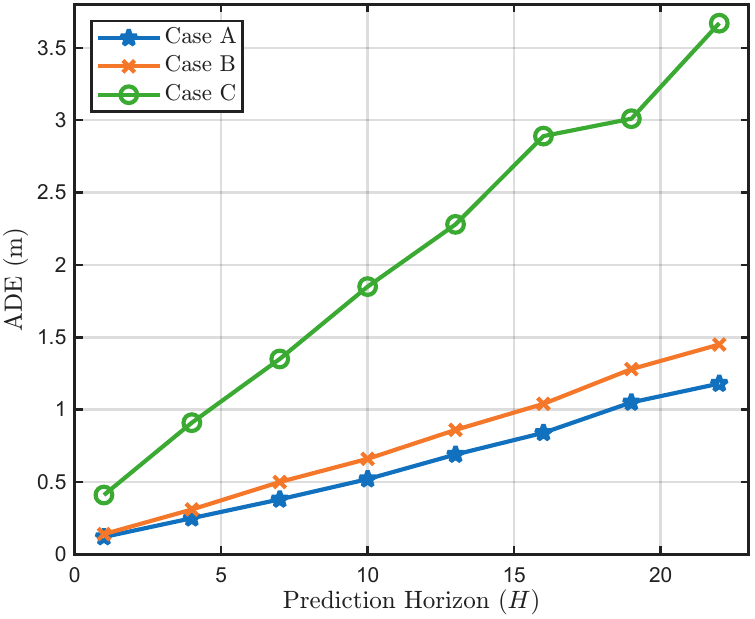}
				\caption{}
				\label{fig:Steps}
			\end{subfigure}
			
			\vspace{0.5em}
			
			\begin{subfigure}{0.32\textwidth}
				\centering
				\includegraphics[width=0.9\linewidth]{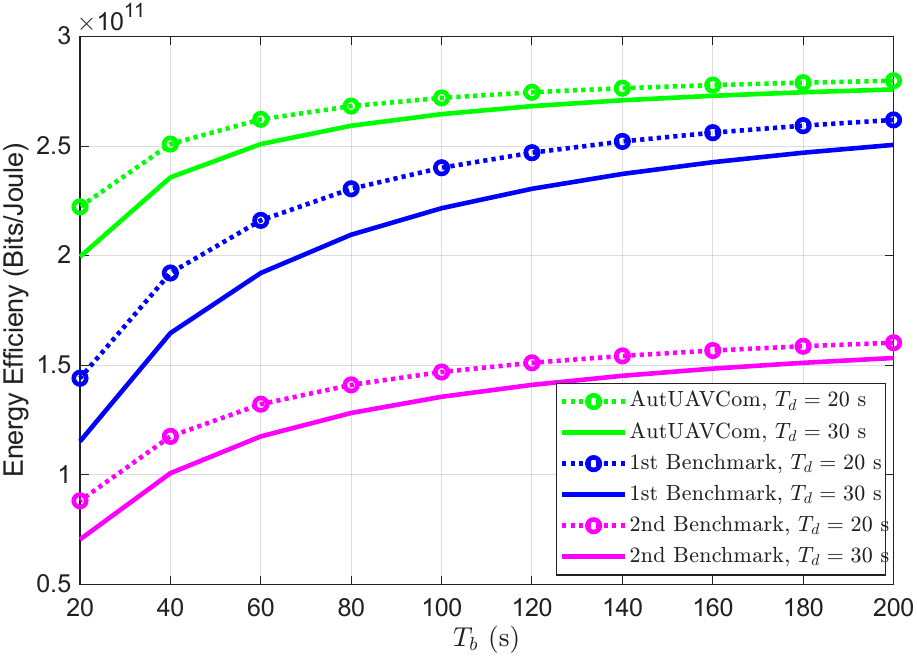}
				\caption{}
				\label{fig:EE}
			\end{subfigure}
			\hfill
			\begin{subfigure}{0.32\textwidth}
				\centering
				\includegraphics[width=0.9\linewidth]{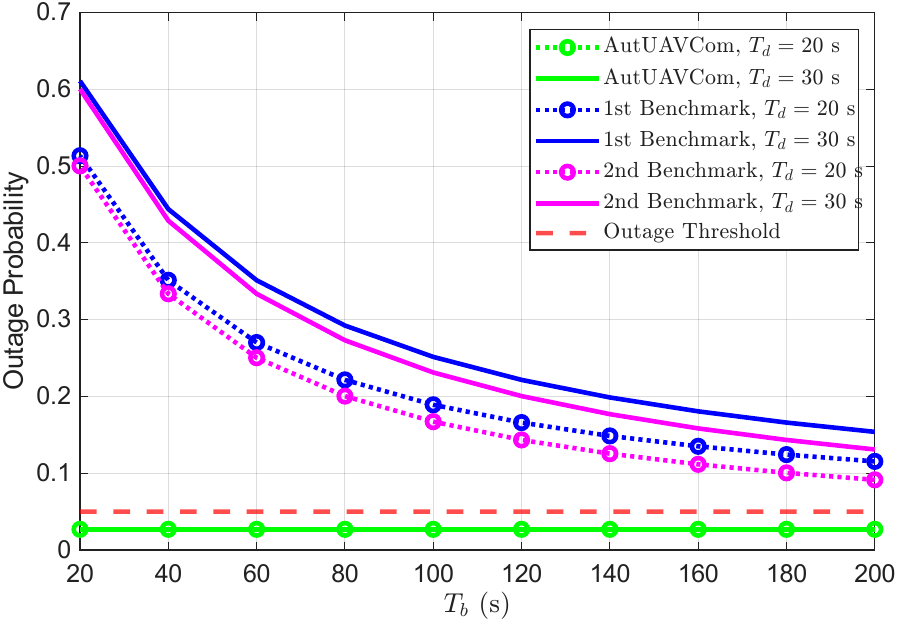}
				\caption{}
				\label{fig:OA}
			\end{subfigure}
			\hfill
			\begin{subfigure}{0.32\textwidth}
				\centering
				\includegraphics[width=0.9\linewidth]{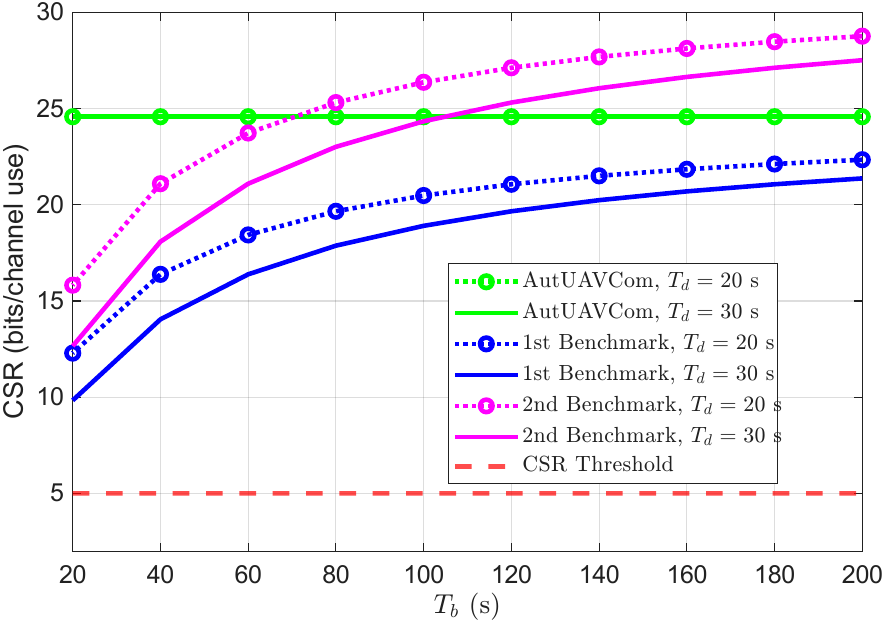}
				\caption{}
				\label{fig:CSR}
			\end{subfigure}
			
			\vspace{-0.5em}
			\caption{Performance evaluation of the proposed AutUAVCom system:
				(a) training errors of the T\&P module,
				(b) ADE histogram,
				(c) ADE versus prediction horizon $H$,
				(d) energy efficiency,
				(e) outage probability, and
				(f) CSR.}
			\label{fig:AutUAVCom_Performance}
			\vspace{-0.4cm}
		\end{figure*}
\section{Numerical Results}
\label{Numerical_Results}
We evaluate the AutUAVCom system against two reactive benchmarks.
The first benchmark solves the same optimization problem in ~\eqref{eq:Pro_Form_MPC}  with $H=1$, making decisions based only on the current observation. The second is a K-means-based benchmark in which the UAV follows the user-group centroid without explicitly optimizing its trajectory. We simulate a 5-user group, across a $200\times100 \text{m}^2$ urban environment with propagation parameters $a=9.61$ and  $b=0.16$. Blockages follow a Poisson process with arrival intervals $T_b \in [20, 200]\text{s}$ and durations $T_d \in \{20, 30\}\text{s}$, while their spatial dimensions follow a log-normal distribution with parameters $(2, 0.5)$ m. The group covers approximately $2200\, \text{m}$ over a $25$-minute mission. The T\&P module is trained offline using an RPGM-GMM dataset with a total trajectory length of $\approx 13{,}019 \text{m}$, an average step length of $\approx 1.3 \text{m}$, and a prediction horizon of $\approx 13 \text{m}$ for $H=10$ and $L=40$.\\
Figs.~\ref{fig:AutUAVCom_Performance}(a)-(c) evaluate the T\&P modules across three cases: (A) noise-free with body orientation, (B) noise-free without body orientation, and (C) noisy (standard deviation of $60$ cm) without body orientation. In Fig.~\ref{fig:AutUAVCom_Performance}(a), the MSE and mean absolute error (MAE) convergence curves over 25 training and validation epochs indicate stable learning for Cases~A and~B. Fig.~\ref{fig:AutUAVCom_Performance}(b) shows the Average Displacement Error (ADE) distributions, where Case~A yields the narrowest and lowest-error spread, whereas Case~C exhibits a broader, right-shifted distribution. As shown in Fig.~\ref{fig:AutUAVCom_Performance}(c), ADE increases monotonically with the prediction horizon $H \in \{1,\dots,22\}$ for all cases; however, the increase is most pronounced for Case~C, indicating greater sensitivity to noise at longer horizons. Overall, these results show that incorporating body orientation improves trajectory prediction accuracy, particularly over longer horizons, whereas noise contamination progressively degrades prediction performance as the prediction horizon increases. Figs.~\ref{fig:AutUAVCom_Performance}(d)-(f) evaluate the communication performance. As shown in Fig.~\ref{fig:AutUAVCom_Performance}(d),  energy efficiency improves as $T_b$ increases and $T_d$ decreases, because more frequent and longer blockages require additional UAV adaptation. AutUAVCom consistently outperforms the benchmarks due to its improved resilience to beam blockages. Meanwhile, the 1st Benchmark consistently outperforms the 2nd Benchmark due to its optimized trajectory. In Fig.~\ref{fig:AutUAVCom_Performance}(e), AutUAVCom maintains a near-target outage probability, whereas the benchmarks fail under severe blockage conditions (small $T_b$, large $T_d$), saturating near $0.1$. However, the 2nd Benchmark performs better than the 1st Benchmark due to the UAV's closer proximity to the user-group centroid. Finally, Fig.~\ref{fig:AutUAVCom_Performance}(f) demonstrates that AutUAVCom maintains a CSR well above the required threshold, while the benchmarks increase rapidly, particularly the 2nd Benchmark due to the UAV's closer proximity to the user-group centroid, and plateau at a higher level as blockages become less severe (large $T_b$, small $T_d$).  Numerically, AutUAVCom significantly outperforms  1st Benchmark, achieving $10\%$ and $19\%$ gains in energy efficiency and CSR, respectively, while reducing the outage probability by $85\%$, when  $T_b=120$ and  $T_d=20$ sec.
\vspace{-0.2cm}
\section{Conclusion}
\label{Conclusion}
\vspace{-0.2cm}
The AutUAVCom system is a data-driven, proactive UAV communication system that fuses onboard sensing with DNN-based trajectory prediction to overcome beam blockages in challenging environments, thereby improving energy efficiency, CSR, and reliability. While AutUAVCom outperforms reactive benchmarks, its performance depends on prediction accuracy, as well as the efficiency of the optimization process. Moreover, AutUAVCom requires a sufficiently large and reliable dataset, which may not always be available in new or unfamiliar environments. In such cases, future work will investigate adaptive or transfer-learning strategies for updating the pretrained model using data collected in new environments, as well as developing a digital twin to reduce the computational and hardware burden onboard UAVs and validating the proposed approach on real-world flight testbeds.
\vspace{-0.6cm}
\bibliographystyle{IEEEtran}
\bibliography{Ref}

@article{Ghazy_2026_UAV,
  title={Enhancing UAV Trajectory and Communications Through Vision-Inertial Tracking},
  author={Ghazy, Abdallah S. and Ammar, Hussein A. and Bayes, James and Chan, Francois},
  journal={arXiv preprint arXiv:2609.03099},
  year={2026}
}

@Article{2014_PathLoss,
  Title                    = {Optimal LAP Altitude for Maximum Coverage},
  Author                   = {Al-Hourani, Akram and Kandeepan, Sithamparanathan and Lardner, Simon},
  Journal                  = {IEEE Wireless Commun. Lett.},
  Year                     = {2014}
}

@Book{MPC_Book_2025,
  Title                    = {Model Predictive Control},
  Author                   = {Camacho, Eduardo F. and Bordons, Carlos and Maestre, Jos{\'e} M.},
  Publisher                = {Springer},
  Year                     = {2025}
}

@Article{Farid_Steve,
  Title                    = {Outage Capacity Optimization for Free-Space Optical Links With Pointing Errors},
  Author                   = {Farid, Ahmed A. and Hranilovic, Steve},
  Journal                  = {J. Lightwave Technol.},
  Year                     = {2007}
}

@Book{Gallager2013,
  Title                    = {Stochastic Processes: Theory for Applications},
  Author                   = {Gallager, Robert G.},
  Publisher                = {Cambridge University Press},
  Year                     = {2013}
}

@Article{Ghazy_UAV_PDFs,
  Title                    = {IRS-Aided Secure Reliable Underwater Acoustic Communications},
  Author                   = {Ghazy, Abdallah S. and Kaddoum, Georges and Satinder, Singh},
  Journal                  = {IEEE Trans. Veh. Technol.},
  Year                     = {2024}
}

@Book{2005_Goldsmith_Wireless,
  Title                    = {Wireless Communications},
  Author                   = {Goldsmith, Andrea},
  Publisher                = {Cambridge Univ. Press},
  Year                     = {2005}
}

@InProceedings{UAV_FullDuplex_2025,
  Title                    = {Online 3D Trajectory and Transmit Power Optimization for Securing UAV-Assisted Full-Duplex Communication Network},
  Author                   = {Huang, Zhiyu and Wang, Yi and Jiang, Jun and Nasir, Ali A. and Sheng, Zhichao and Yang, Xue-Xia},
  Booktitle                = {IEEE VTC-Fall},
  Year                     = {2025}
}

@Article{WNCS_2026,
  Title                    = {MPC-Based Path Tracking and Power Allocation for WNCS: An Outage Probability Perspective},
  Author                   = {Jin, Haijia and Wu, Jun and Yuan, Weijie},
  Journal                  = {IEEE Wireless Commun. Lett.},
  Year                     = {2026}
}

@InProceedings{body_orientation_2009,
  Title                    = {Tracking of humans and estimation of body/head orientation from top-view single camera for visual focus of attention analysis},
  Author                   = {Ozturk, Ovgu and Yamasaki, Toshihiko and Aizawa, Kiyoharu},
  Booktitle                = {IEEE ICCV Workshops},
  Year                     = {2009}
}

@Article{3D_Securing_2020,
  Title                    = {Securing UAV Communication in the Presence of Stationary or Mobile Eavesdroppers via Online 3D Trajectory Planning},
  Author                   = {Savkin, Andrey V. and Huang, Hailong and Ni, Wei},
  Journal                  = {IEEE Wireless Commun. Lett.},
  Year                     = {2020}
}

@Article{MPC_Security_2025,
  Title                    = {Outage-Aware Online Prediction Control for Securing UAV-Aided Communication},
  Author                   = {Sheng, Zhichao and Fu, Huabo and Huang, Zhiyu and Nasir, Ali A. and Wu, Qingqing and Zeng, Dan},
  Journal                  = {IEEE Trans. Veh. Technol.},
  Year                     = {2025}
}

@Article{2013_Hovering_PDF,
  Title                    = {Analysis of Wind Effect on Drone Relay Communications},
  Author                   = {Tajima, Yuya and Hiraguri, Takefumi and Matsuda, Takahiro and Imai, Tetsuro and Hirokawa, Jiro and Shimizu, Hiroyuki and Kimura, Tomotaka and Maruta, Kazuki},
  Journal                  = {Drones},
  Year                     = {2023}
}

@Article{MPC_Solar_2021,
  Title                    = {MPC-Based UAV Navigation for Simultaneous Solar-Energy Harvesting and Two-Way Communications},
  Author                   = {Tuan, Hoang Duong and Nasir, Ali Arshad and Savkin, Andrey V. and Poor, H. Vincent and Dutkiewicz, Eryk},
  Journal                  = {IEEE J. Sel. Areas Commun.},
  Year                     = {2021}
}

@Article{UAV_Survey_2019,
  Title                    = {Accessing From the Sky: A Tutorial on UAV Communications for 5G and Beyond},
  Author                   = {Zeng, Yongs and Wu, Qingqing and Zhang, Rui},
  Journal                  = {IEEE},
  Year                     = {2019}
}

@Article{2019_Pwer_Hovering_Model,
  Title                    = {Energy Minimization for Wireless Communication With Rotary-Wing UAV},
  Author                   = {Zeng, Yong and Xu, Jie and Zhang, Rui},
  Journal                  = {IEEE Tr. Wi. Com.},
  Year                     = {2019}
}

\end{document}